\documentclass[floatfix,pre,twocolumn,showpacs,preprintnumbers,amsmath,amssymb]{revtex4}
\usepackage{graphicx}

\usepackage{hyperref}

\newcommand{\units}[2]{\ensuremath{#1\,\mathrm{#2}}}
\newcommand{\runits}[3]{\ensuremath{#1\mbox{--}#2\,\mathrm{#3}}}

\newcommand{\wheel}{\ensuremath{\mathrm{wheel}}}
\newcommand{\Fr}{\ensuremath{\mathrm{Fr}}}
\newcommand{\Frc}{\ensuremath{\mathrm{Fr_c}}}

\newcommand{\Frcw}{\ensuremath{\mathrm{Fr_c^{wheel}}}}
\newcommand{\speed}[1]{\units{#1}{m\,s^{-1}}}
\newcommand{\ie}{\textit{i.e.}}
\newcommand{\vs}{\textit{vs.}}

\def\gsim{\mathrel{\rlap{\lower4pt\hbox{\hskip1pt$\sim$}}
    \raise1pt\hbox{$>$}}}                

\begin{document}

\title{Scaling and dynamics of washboard road}

\author{Anne-Florence Bitbol$^{1,2}$, Nicolas Taberlet$^{1}$,
 Stephen W.\ Morris$^3$, Jim N.\ McElwaine$^2$}

\affiliation{
 $^{1}$ Universit\'e de Lyon, \'Ecole Normale Sup\'erieure de Lyon, Laboratoire de Physique, 
 46 all\'ee d'Italie, 69007 Lyon, France\\
 $^{2}$DAMTP, University of Cambridge, Wilberforce Rd., CB3 0WA Cambridge, U.K.\\
 $^{3}$Department of Physics, University of Toronto, 60 St. George
 St., Toronto, Ontario, Canada, M5S 1A7}

\date{\today}

\begin{abstract} 
  Granular surfaces subjected to forces due to rolling wheels develop
  ripples above a critical speed. The resulting pattern, known as
  \emph{washboard} or \emph{corrugated} road, is common on dry,
  unpaved roads. We investigated this phenomenon theoretically and
  experimentally, using laboratory-scale apparatus and beds of dry
  sand. A thick layer of sand on a circular track was forced by a
  rolling wheel on an arm whose weight and moment of inertia could be
  varied. We compared the ripples made by the rolling wheel to those
  made using a simple inclined plow blade. We investigated the
  dependence of the critical speed on various parameters, and describe
  a scaling argument which leads to a dimensionless ratio, analogous
  to the hydrodynamic Froude number, which controls the instability.
  This represents the crossover between conservative, dynamic forces
  and dissipative, static forces. Above onset, wheel-driven ripples
  move in the direction of motion of the wheel, but plow-driven
  ripples move in the reverse direction for a narrow range of Froude
  numbers.
 
\end{abstract}

\maketitle

The spontaneous rippling of gravel roads after the passage of many
vehicles is annoyingly familiar to drivers of unpaved roads around the
world. Avoiding or mitigating the rippling effect, known as
\emph{washboard} or \emph{corrugated} road, is a significant
engineering challenge~\cite{heath80:washreview, mather62:washboard1,
  shoop06:washboard, riley73:washboard, riley71:washboard,
  grau91:washboard, misoi89:washboard1}. We have argued
recently~\cite{taberlet07:washboard} that the rippling can be regarded
as a type of nonlinear pattern-forming
instability~\cite{cross04:pattern} of the flat
road~\cite{mays00:washboard,both01:washboard}. This nonlinear physics
point of view is in many ways complementary to the engineering one,
and leads to new insights. We present here an experimental and
theoretical study of the region near the onset of the instability,
which appears only above a critical threshold speed. We focus on how
this critical speed scales with various parameters. We also survey
some of the complex dynamics that occurs above
onset~\cite{mays00:washboard}.

Washboard road is commonly, and incorrectly, assumed to require the
forced oscillation of the suspension of the
vehicle~\cite{misoi89:washboard1}. Although a suspension system does
change the quantitative details, ripple frequencies are generally
quite far from the free resonant frequencies of the
suspension~\cite{mather62:washboard1,misoi89:washboard1}, and ripples
can appear even with no suspension
present~\cite{taberlet07:washboard}.  Here, we consider a simplified
experimental system in which a wheel merely returns to the road
surface under gravity. We also study the even simpler case of a
non-rotating, angled ``plow'' blade in place of the rolling wheel. In
both cases, well above onset, we find that the ripples travel down the
road in the direction of driving and that the wheel or plow is thrown
free of the roadbed between ripples. In the case of the plow, there
also exists a range just above onset where the blade remains in
contact with the surface and the ripples travel in the reverse
direction, against the direction of driving.

The rippling instability is quite robust, and has analogs in other
similar effects on granular surfaces, such as wind~\cite{bagnold1941}
and water-driven~\cite{engelund82:ripples} ripple patterns. Other
related phenomena are, for example, periodic wear patterns on railroad
tracks~\cite{sato02:rails}, mogul formation on ski
slopes~\cite{egger02:moguls}, skipping stones over
water~\cite{clanet04:skipping} and certain failure modes of computer
hard disks~\cite{dai04:washboard,pit01:mogul}. What all these
phenomena have in common is a repeated, nonlinear interaction between
a moving source of lateral and vertical contact stresses acting on a
deformable or erodible surface. The interaction includes the case when
the source of stress, which both deforms and responds to the shape of
the surface, completely leaves the surface so that the stresses go to
zero. We propose a very general picture of the key mechanism of the
washboard instability in terms of a dimensionless group, analogous to
the hydrodynamic Froude number, which we show controls the scaling of
the onset speed. This analysis may be broadly generalizable to the
diverse situations cited above.

Previous experimental work and soft sphere molecular dynamics
simulations~\cite{taberlet07:washboard} have shown that neither
compaction nor size segregation processes in the granular bed are
essential for the instability, contrary to some
theories~\cite{both01:washboard}. Engineering
studies~\cite{heath80:washreview,mather62:washboard1,shoop06:washboard,riley73:washboard,riley71:washboard,grau91:washboard}
have mainly concentrated on optimizing the response of a suspension
system to fully developed ripples and have not typically considered
the threshold of the instability. Using our approach, it may be
possible to suppress or delay the onset of this threshold by using
appropriate control strategies based on the nonlinear dynamics of the
coupled wheel-roadbed system, which behaves in some respects like an
impact oscillator~\cite{chin94:impact,bishop94:impact}.

\begin{figure*}[htbp]
  \parbox{0.9\textwidth}{
    \mbox{}%
    \hfill%
    \includegraphics[height=55mm]{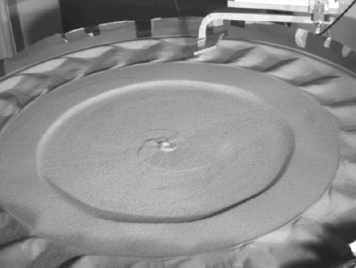}
    \hfill%
    \includegraphics[height=55mm]{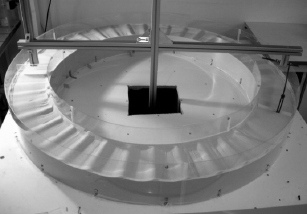}
    \hfill
    \mbox{}%
  } 
  \caption{The two versions of the experimental
    apparatus. The left photo shows the setup in Cambridge, where the
    table carrying the sand roadbed rotates beneath an arm holding a
    wheel or plow, which is stationary in the lab frame. The right
    photo shows the setup in Lyon, where the roadbed is stationary
    beneath an arm that rotates. In both experiments the moving parts
    rotate counterclockwise, as seen from above.}
  \label{fig:exp_photo}
\end{figure*}

This paper is organized as follows; in the section~\ref{sec:method},
we describe the experimental apparatus. This is followed in
section~\ref{sec:results} by a qualitative discussion of the rippling
phenomena that are observed for both wheels and plow blades. In
section~\ref{sec:theory}, we consider the theory of the basic scaling
of the onset speed, which we then examine experimentally in
section~\ref{sec:scaling}. Next, in section~\ref{sec:backward} we
discuss the regime of backward traveling ripples and in
section~\ref{sec:zoo} we describe some of the rich variety of ripple
states that appear above onset. Finally, section~\ref{sec:conclusion}
is a brief conclusion.

\section{Experimental Method}
\label{sec:method}

Two versions of the experimental apparatus were used in this study, as
shown in Fig.~\ref{fig:exp_photo}. The first was essentially that of
Ref.~\cite{taberlet07:washboard} and consisted of a \units{1}{m}
diameter rotating table, with a maximum rotation speed of
\units{0.8}{Hz}, holding a \units{200}{mm} deep layer of rough sand,
producing a moving roadbed.  This passed under an arm which could hold
either a rolling wheel or a plow blade which was stationary in the lab
frame. This arrangement has some advantages for visualization and also
the large moment of inertia of the table helps maintain a constant
speed. In the second version of the apparatus, the arm holding the
wheel or plow hangs from a rotating axle over a \units{1.80}{m}
diameter roadbed which is stationary in the lab frame. This version is
larger and can reach higher speeds, with a maximum axle rotation
frequency up to \units{1}{Hz}. In both versions, we used rough sand
with a grain diameter of \units{300\pm100}{\mu m}. We have previously
shown~\cite{taberlet07:washboard} that the shape and size of the
grains is unimportant to the rippling phenomenon. The density of the
grains, which was not varied in the present study, is however a
crucial parameter, as discussed in section~\ref{sec:theory} below.

Fig.~\ref{fig:exp_schematic} shows the arm with the wheel and plow
blade configuration in detail. A light, 330\,mm long arm pivoted
around a point of support. An optional counterweight could be added to
the end opposite the wheel or plow. The weight and transverse width of
the wheel or plow can be varied, and, in the case of the plow, the
angle of attack $\alpha$ was also adjustable.  The wheel used was
smaller and lighter than that described in
Ref.~\cite{taberlet07:washboard}, having a diameter of \units{65}{mm}
and a typical width of \units{25}{mm}. The plow blade was typically
\units{120}{mm} wide and inclined at an angle of $\alpha=40^{\circ}$.

The masses of the wheel or plow and counterweight could be varied
between \units{0.045}{kg} and \units{3}{kg}. As discussed in
section~\ref{sec:theory} below, these arrangements are dynamically
equivalent to a physical pendulum, as shown in
Fig.~\ref{fig:exp_schematic}c. Since the angle of the arm $\theta$ was
small, the physical pendulum was always nearly horizontal.  Thus, the
wheel or plow behaved as if it were a freely falling object and the
natural frequency of the pendulum played no role. No spring or dashpot
was used, so this arrangement constitutes a minimal suspension having
no resonant or natural frequencies, and only rather weak damping.

The wheel rolled freely and no torques were applied to it other than
that produced by contact with the surface. The plow was simply dragged
over the surface. The speeds were inferred from measurements of the
rotation rate of the table and spanned the range
\runits{0}{6}{m\,s^{-1}}.  A magnetic angle
sensor~\cite{magnetic_sensor} was attached to the pivot supporting the
arm, and its angle $\theta$ was digitized and used to find the
vertical position $z$ of the wheel or plow blade.  $z$ was measured
with a vertical accuracy of 0.2\,mm at a rate of 2000\,Hz. The
position data was acquired continuously along with data from an optical
encoder that generated a pulse each table rotation.

\begin{figure}[htbp]
 \includegraphics[width=70mm]{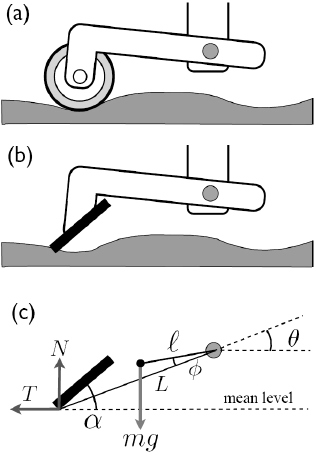}
 \caption{Schematic views of (a) the wheel and (b) plow blade. (c)
   shows the geometry of the equivalent physical pendulum. The plow is
   rigidly fixed to the arm and makes an angle of attack $\alpha$ with
   respect to the horizontal.}
  \label{fig:exp_schematic}
\end{figure}

The dynamics of the ripple pattern evolved rather slowly, especially
near the critical speed $v_c$ for the onset of the instability. It was
thus necessary to increase and decrease the speed over periods of many
hours to achieve quasi-static, steady state patterns.  In some runs,
the speed was ramped up quickly from a standing start, beginning with
a flat sand bed.  This protocol is analogous to making a rapid
``quench'' into the rippled state. In other runs, a small perturbation
was made in the bed to act as a nucleation site for ripples.

\section{Qualitative Results}
\label{sec:results}

In this section, we discuss some of the phenomena observed, before
focusing on the scaling of the critical speed for the onset of
rippling in section~\ref{sec:scaling}. Initially, we tried preparing
the roadbed by running a comb through it to make a loose, flat,
granular assembly. However, if the experiment is run continuously over
a period of around a week ($\approx$ 100\,000 rotations), the bed
gradually compacts and the thresholds and amplitudes slowly change. We
therefore used a different protocol where the sand was disturbed as
little a possible. To prepare a flat bed, the experiment was run at a
speed below the onset for many rotations. The results were then
repeatable.

\begin{figure}[htbp]
  \includegraphics{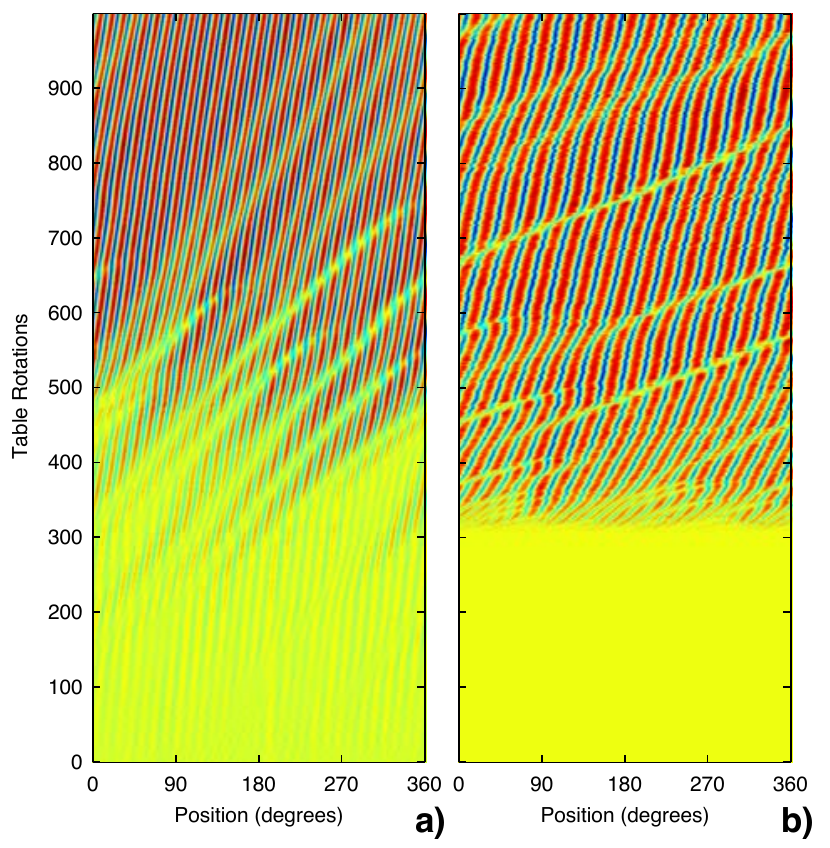}
  \caption{(Color online) Typical space-time plots showing the
    development of ripples from a flat bed, with red indicating a
    higher arm position and blue a lower. Both parts show 1000
    rotations and have the same scale. The driving direction is to the
    right.  a) A free wheel just above the critical velocity. b) The
    plow, with $v$ below the critical velocity for the first 300
    rotations and well above critical for the next 700 rotations.}
  \label{fig:init}
\end{figure}

Ripples driven by either the wheel or the plow only appear above a
well-defined critical speed $v_c$. Below this speed, perturbations
made in the roadbed are observed to be smoothed away by the action of
the wheel or plow. For speeds near $v_c$, the growth rate of
perturbations is nearly zero, and it can take many thousands of
rotations of the experiment before ripples build up.  The quasi-steady
state ripples which develop after the initial transient have a very
small amplitude just above $v_c$. For all cases, speeds well above
$v_c$ result in regimes where the wheel or plow is thrown completely
off the roadbed near the crest of a ripple, before landing again on
the front face of the next ripple. The resulting large amplitude
ripples are strongly asymmetric, with a gradual rise on their front
faces, and an abrupt slip-face on the back side. Nearer to $v_c$, the
ripples are more symmetrical.

\begin{figure}[htbp]
  \includegraphics{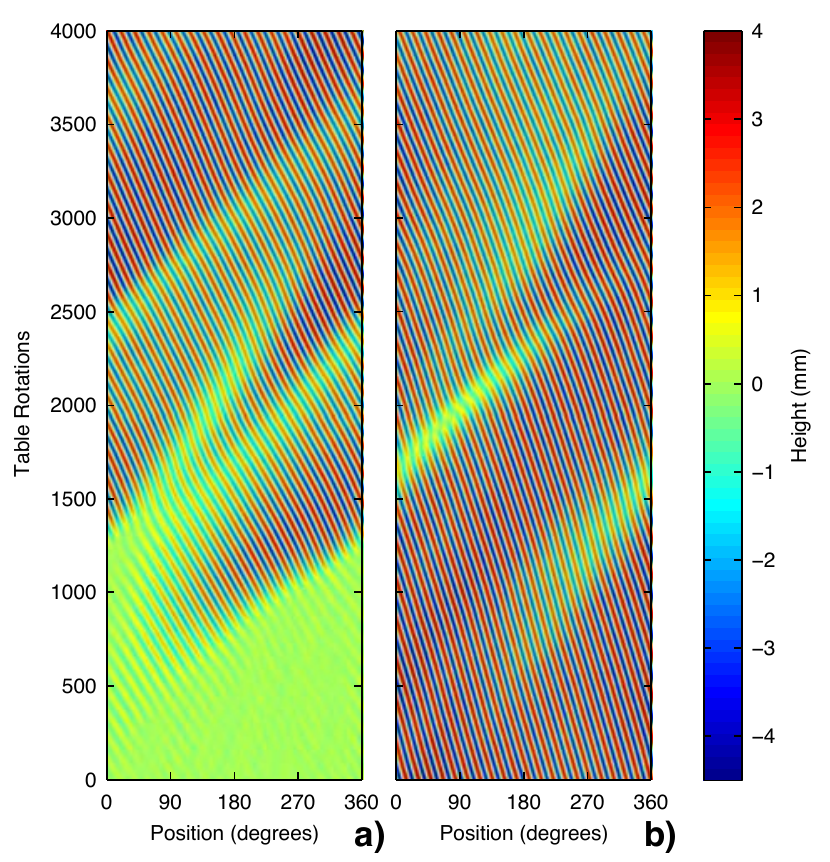}
  \caption{(Color online) Parts a and b show different segments of a
    ten day experiment during which the speed was decreased in steps
    of \speed{0.0021} every 500 rotations. The driving direction is to
    the right.  In part a, the speed is increasing from \speed{0.684}
    to \speed{0.691} (\Fr ~2.04--2.09), triggering a transition from
    flat bed to backward traveling ripples. In b, The speed is
    decreasing from \speed{0.694} to \speed{0.687} (\Fr ~2.10--2.06),
    triggering a transition from 21 to 20 ripples.  Here
    $M_G=\units{0.0785}{kg}$, $w=\units{0.120}{m}$,
    $\rho=\units{1201}{kg\,m^3}$, and the angle of attack $\alpha$ was
    \units{36.8}{^\circ}}
  \label{fig:backwards}
\end{figure}
The plow is observed to remain in contact with the surface at all
times near $v_c$. The ripples move in the direction of driving for
wheels, but may move in the reverse direction for plows, if $v$ is
close to critical.  At larger speeds, ripples move in the direction of
driving in both cases.  We discuss this curious reversing behavior in
section~\ref{sec:backward} below. Fig.~\ref{fig:init} shows some
representative space-time plots of ripple evolution after a quench
into the strongly unstable regime. During the transient, numerous
ripple creation and annihilation events are observed, resulting in
adjustments of the pattern which tend to propagate in the direction of
driving. Fig.~\ref{fig:backwards}a shows a quench close to the
threshold with the plow showing backward ripples. Eventually, a
quasi-steady state of traveling ripples is established.  For
sufficiently deep quenches, the pattern can display multi-modal
behavior or persistent time dependence, as discussed in
section~\ref{sec:backward}. The final, saturated amplitude of the
ripples increases with the driving speed near the threshold, but shows
non-monotonic behaviour far above.

It is easy to observe experimentally that $v_c$ depends on the mass of
the wheel or plow, and the moment of inertia of the arm. By changing
the granular material, it was apparent~\cite{taberlet07:washboard}
that only the density of the roadbed material matters, and not the
size or shape of the grains. These dependencies can be accounted for by
the scaling analysis discussed in the next section.

\section{Scaling Theory}
\label{sec:theory}

In this section, we consider the dynamics of the arm assembly and its
interaction with the sand bed. We seek to identify the essential
physical parameters that govern the scaling of the critical onset
speed $v_c$.  To simplify the problem, we consider a strictly
two-dimensional model in which the transverse width of the wheel or
plow is assumed to be large compared to its penetration into the bed.
This assumption was exactly true for the soft sphere molecular
dynamics simulations discussed in Ref.~\cite{taberlet07:washboard},
which qualitatively reproduced all of the essential features of the
experiment.  The validity of this simplifying assumption for the real
experiment will be discussed in section~\ref{sec:scaling}.

The geometry of the arm and plow assembly is shown in
Fig.~\ref{fig:exp_schematic}c. The position of the assembly is
completely specified by three points: the contact point with the bed,
the pivot position and the location of the center of mass.  These
three positions define two intrinsic lengths, the distance from the
pivot point to the center of mass $\ell$, and the distance from the
pivot point to the contact point with the bed $L$.  The angle $\phi$
between these lines is constant.  The angle $\alpha-\theta$ between
the plow and the line to the pivot point is also fixed, whereas
$\theta$ will vary as the bed deforms.

Dynamically, the system can be described as a forced physical pendulum
of length $\ell$, mass $m$ and moment of inertia about the center of
mass $I$.  The latter two quantities define another length
$\sqrt{I/m}$.  There is one additional length, the transverse width of
the plow or wheel $w$.  Thus, to describe the whole assembly requires
three angles, $\alpha$, $\theta$ and $\phi$ (which are of course
non-dimensional), four lengths $\ell$, $L$, $\sqrt{I/m}$ and $w$, the
mass $m$ and the acceleration due to gravity $g$.  The properties of
the bed involve only one additional parameter, the sand density
$\rho$, as discussed below.

The gravitational force $mg$ acts through the center of mass and
exerts a torque $m g \ell \cos (\theta-\phi)$ about the pivot point.
The force of the bed on the arm is assumed to act entirely at the
contact point.  It exerts a torque about the pivot point of $ -
L(N\cos\theta + T\sin\theta)$, where $N$ is the vertical (normal)
force from the bed and $T$ the horizontal (tangential) force.  Since
the moment of inertia about the pivot point is $I+m\ell^2$,
conservation of angular momentum gives
\begin{equation}\label{eq:z0}
 (I+m\ell^2) \ddot \theta = m g\ell \cos (\phi-\theta) - L(N\cos\theta + T\sin\theta).
\end{equation}

The vertical height of the contact point relative to the pivot point
is $z=-L\sin\theta$.  To determine how the onset speed $v_c$ scales,
we consider small displacements about a flat bed, so that $\theta$ is
small. This is sufficient, since the transition to rippling occurs at
zero amplitude.  We treat $\theta$ as constant except in the $\ddot
\theta$ term where $\ddot z=-L \ddot \theta$.  After dividing
Eqn.~\ref{eq:z0} by $L$ and substituting for $z$ we get approximately
\begin{equation}\label{eq:z1}
  \frac{M_I}{\cos\theta}\ddot z = (N\cos\theta + T\sin\theta) -  M_G~g \cos (\theta-\phi),
\end{equation}
where $M_I=(I+m\ell ^2)/L^2$ is the effective inertial mass and
$M_G=m\ell/L$ is the effective gravitational mass.

$M_I$ and $M_G$ can be changed independently by adjusting $\ell$, $L$
and $m$. They are equal if all the mass is located at the end of the
arm. If the angular dependencies are absorbed into the definitions,
that is with $M_I'= {M_I}/{\cos\theta}$, $M_G'=\cos (\theta-\phi)M_G$,
and $N'=N\cos\theta + T\sin\theta$, then for small displacements ({\it
  i.e.} small changes in $\theta$) we have the equation of motion
\begin{equation}\label{eq:z3}
  M_I'\ddot z = N' - M_G' ~g,
\end{equation}
which is simply that of a falling mass interacting with a surface via
a normal force $N'$.  This equation holds for small $\theta$, but for
larger $\theta$, the coupling of the resistive force $T$ to the
vertical motion becomes important. Our experiments concentrated on
small $\theta$, but in subsection~\ref{ssec:angle}, we discuss the
results of experiments where this condition is relaxed.

The mechanics of the granular road bed are not so straightforward to
model as those of the arm.  We aim for a minimal description that
leads to the correct scaling of the fully coupled problem.  Much more
detailed analyzes of the rheology of the granular material would be
required for a very realistic description~\cite{shoop06:washboard}.

We consider the case where the depth of penetration into the bed is
small compared to the width of the object, so that we may assume that
the induced flow in the sand will be essentially two-dimensional, and
that edge effects are small. This assumption is probably violated in
our experiments with the narrowest wheels and plows.

The sand grains were dry and sufficiently large so that they were
non-cohesive.  Previous work~\cite{taberlet07:washboard} has shown
that changing the grain size does not change the wavelength, amplitude
or critical velocity of the ripples.  Therefore, the granular medium
can be described by only its three-dimensional density $\rho$ and
possibly by some other non-dimensional parameters.  The shallowness
assumption and two-dimensionality means that the density $\rho$ of the
sand and the width $w$ will only occur in the form of a
two-dimensional density $\rho w$, and that the width should have no
further role. It may be possible to allow for some three dimensional
effects by allowing a virtual origin correction in the width, though
we have not tested this.

To arrive at a plausible scaling, we examine the balance of forces for
a plow or wheel moving over and penetrating the sand surface. Forces
between the bed and the plow or wheel arise in two distinct ways which
scale differently with $v$. One of these is almost entirely
dissipative and the other one is almost entirely conservative. We postulate that
the onset of washboard ripples occurs when the relative sizes of these
forces cross over as a function of $v$.  We consider the geometrically
simpler case of the plow first.

Static, buoyancy or frictional forces must be proportional to $A \rho
w g$, where $A$ is the displaced area of granular material. For a plow
penetrating a distance $h=z_0-z$, where $z_0$ is the free surface
position, they will be proportional to $h^2 \rho w g$.  These forces
will be almost entirely dissipative. On the other hand, dynamic
forces, those arising from the inertia of the sand, will be
proportional to $h \rho w v^2 $.  These forces are mostly
conservative. We neglect the vertical velocity $\dot h$ as small
compared to $v$. Ignoring constants and angular dependence, we
separately balance these two penetration forces against the effective
weight of the arm $M_G g$.  This gives the static and dynamic
equilibrium penetration depths
\begin{eqnarray}
  h_s &=& \sqrt{\frac{M_G}{ \rho w}},\\
  h_d &=& \frac{M_G g}{v^2 \rho w}. 
\end{eqnarray}
As $v$ increases, the dynamic forces increase and the dynamic
penetration depth $h_d$ needed to support the weight decreases,
whereas the static forces are unaffected. Since the static forces are
almost completely dissipative and the dynamic forces almost completely
conservative, the switch between the two that occurs as $v$ increases
explains the transition from a highly dissipative furrow-plowing mode
to one in which the plow spends some of the time in non-dissipative,
ballistic flight.  For low speeds $v$, the normal forces between the
plow and the sand do only dissipated work against friction, while for
larger $v$, the normal force is sufficient to throw the plow off the
surface and into free flight.  Ripples are then formed and maintained
by the periodic dissipative collisions with the surface. The onset of
washboard ripples in this picture is thus somewhat analogous to the
onset of skipping for a stone projected over a water
surface~\cite{clanet04:skipping}.  A key difference, of course, is
that the granular ripple pattern is built up by many passes of the
plow, each of which contributes only a small, but persistent,
deformation of the surface.  The important non-linearity in the problem
is the highly nonlinear nature of the normal force, which abruptly
goes to zero when the plow leaves the surface, as is the case for
impact oscillators~\cite{chin94:impact,bishop94:impact}.

The dimensionless ratio of the two penetration depths,
\begin{equation}
  \Fr=\frac{h_s}{h_d} = \frac{v^2}{g}\sqrt{\frac{\rho w}{M_G}},
 \label{Fr_def}
\end{equation}
is analogous to the Froude number used in similar hydrodynamic
arguments~\cite{clanet04:skipping}.  We
propose~\cite{taberlet07:washboard} that this dimensionless group
controls the onset of the instability, with ripples setting in above a
critical, threshold value of $\Fr = \Frc$.  Our scaling argument does
not furnish any estimate of the value of $\Frc$, beyond the general
expectation that $\Frc \sim O(1)$.  We establish experimental values
of $\Frc$ in section~\ref{sec:scaling}, below.  We will also show
cases where the plow does not actually leave the surface, but still
forms backward traveling ripples for $\Fr \gsim \Frc $.

We have physically motivated the identification of the Froude number
in terms of a crossover between dissipative and conservative forces~\cite{froude},
but it can also be deduced from straightforward dimensional analysis.
If the only relevant dimensional quantities of the arm that determine
the transition are its effective gravitational mass $M_G$ and its
width $w$, and if we assume the two-dimensional model such that $w$
and $\rho$ can only appear as the product $\rho w$, then $\Fr$ is the
only dimensionless group that can be formed from a combination of
$M_G$, $\rho w$, $v$ and $g$.

The critical Froude number $\Frc$ will also depend on $\theta$,
$\alpha$, $\phi$ and possibly on other dimensionless properties of the
bed. If these are held constant, Eqn.~\ref{Fr_def} implies that the
critical speed $v_c$ is given by
\begin{equation}
  v_c =\left(\Frc\, g\right)^{1/2} \left(\frac{M_G}{\rho w}\right)^{1/4}.
 \label{vc_def}
\end{equation}
In subsection~\ref{ssec:mass}, we will show experimentally that $v_c$, for
the plow, does indeed exhibit an $M_G^{1/4}$ scaling.  In
Sections~\ref{ssec:width} and~\ref{ssec:angle}, we consider the
dependence of $v_c$ on $w$, $\alpha$ and $\theta$.

The wheel is more complicated than the plow in some ways as there is
an additional length scale, the wheel radius $R$. The displaced
volume for a cylinder length $w$ which penetrates to depth $h$ is
\begin{eqnarray}
  && w\bigg[R^2 \cos^{-1}(1-h/R)-(R-h)\sqrt{2Rh-h^2}\bigg] \nonumber \\
  &=& \frac{4}{3}w h \sqrt{2Rh}~\bigg[1 +O({h}/{R})\bigg]. 
\end{eqnarray}
Thus the scaling is now $A \propto h^{3/2}$ instead of $A \propto
h^2$, so that the static depth is modified. For the wheel
\begin{equation}\label{eq:hs}
  h_s^{\wheel} = R^{-1/3}\left(\frac{M_G}{\rho w}\right)^{2/3}.
\end{equation}
The dynamic depth is the same thus the scaling of the Froude number is
changed to
\begin{equation}
  \Fr^{\wheel}=\frac{h_s^{\wheel}}{h_d} = \frac{v^2}{g}\left(\frac{\rho w}{M_G R}\right)^{1/3}.
\end{equation}
Thus, we predict the following relation for the critical speed
\begin{equation}
 v_c^{wheel} =\sqrt{\Frcw} \left(\frac{g^3 M_G R}{\rho w}\right)^{1/6},
 \label{vc_wheel}
\end{equation}
where numerical coefficients have been ignored.

To experimentally determine the relevant parameters of the model, it
suffices to determine the three constant quantities that appear in
Eqn.~\ref{eq:z1}, namely $M_I$, $M_G$ and $\phi$.  We supported the
arm on a balance, so that the bottom of the plow was exactly level
with the pivot point and $\theta=0$.  Then the force measured by the
balance was $N=M_G~g \cos\phi$.  The angle $\phi$ was determined by
suspending the arm from the pivot point and measuring $\theta$. Then
$M_G~g \cos (\theta-\phi)=0$ and $\phi=\pm\pi/2+\theta$.
Eqn.~\ref{eq:z1} for the suspended arm is simply the equation of a
free physical pendulum,
\begin{equation}
 M_IL~\ddot \theta = M_G~g \cos (\phi-\theta),
\end{equation}
which, for small amplitudes, oscillates with angular frequency $\omega
= \sqrt{{M_G\,g}/{M_IL}}$. The length $L$ was directly measured with a
ruler, while $\omega$ was measured by triggering small oscillations,
and then measuring $\theta$ using the magnetic sensor.  In this way,
all the parameters in Eqn.~\ref{eq:z1} were determined.  All these
quantities vary slightly with the angle of attack $\alpha$ of the
plow, and so were measured for a range of $\alpha$. The plows of
different width were carefully chosen so as not to alter any of the
properties.

\section{Scaling of the threshold speed}
\label{sec:scaling}

To test the scaling theory outlined in the previous section, we
measured the critical speed $v_c$ for a variety of plow and wheel
configurations. The transition speed is most accurately determined by
running the system at a constant speed above $v_c$ until a steady
state has developed. The speed is then decreased by steps of
\speed{0.0016} and kept at each speed for 1000\, rotations. This is
more accurate and quicker than starting with an initial perturbation
or flat bed and observing the subsequent behaviour, as a very large
number of rotations are necessary for the ripples to develop near
$v_c$.  We also recorded the long-term dynamical states for a range of
speeds from below to well above onset.

As discussed in the following subsections, we added various masses and
counterweights to the arm, which has the effect of varying $M_I$ and
$M_G$, as well as varying the width of the plow $w$, its angle of
attack $\alpha$ and the mean support angle $\theta$.

\subsection{Variation with gravitational mass $M_G$}
\label{ssec:mass}

\begin{figure}[htbp]
  \includegraphics[width=95mm,clip]{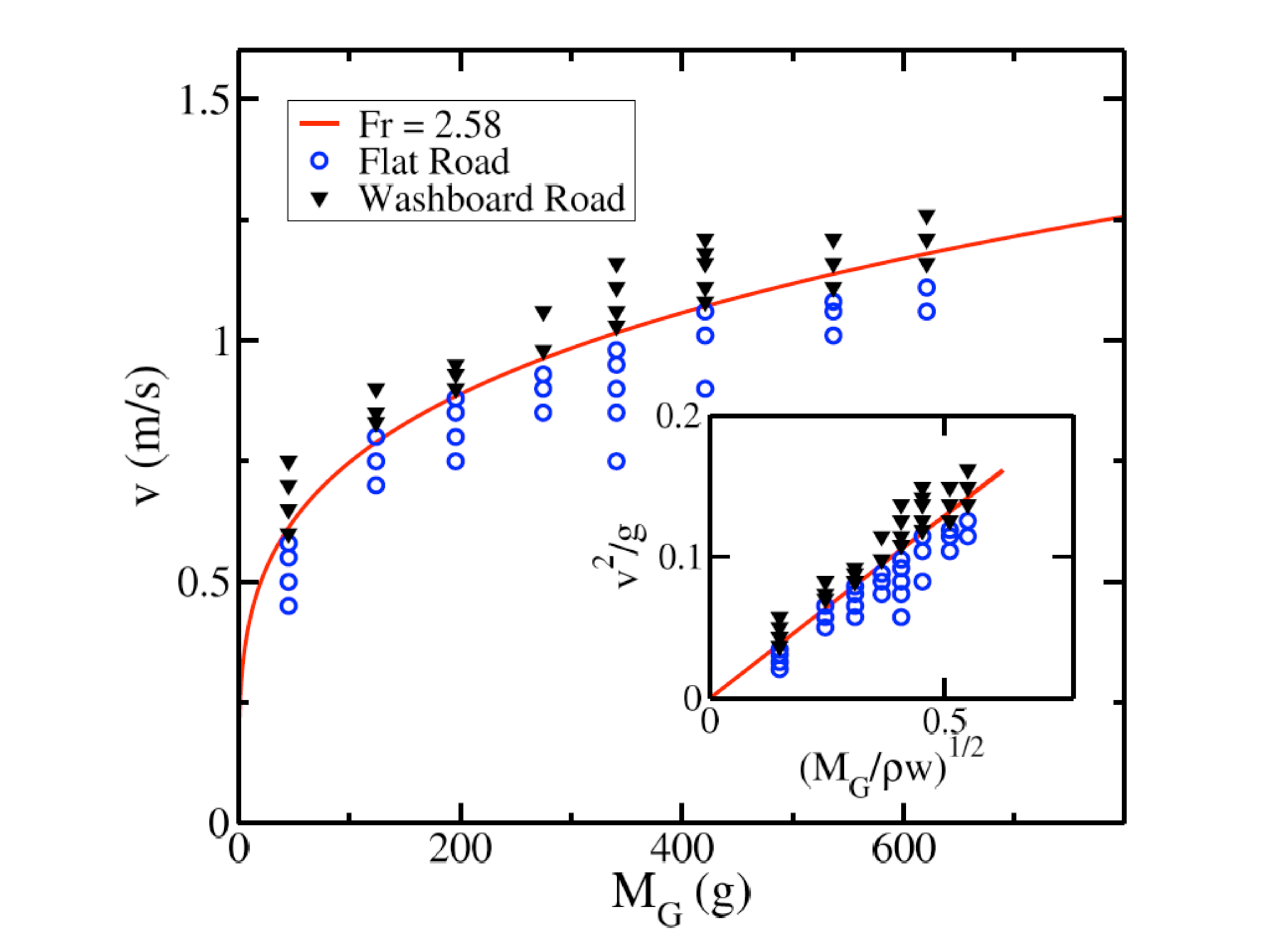}
  \caption{(Color online) The domain of existence of the ripple
    pattern for the plow, as a function of the speed $v$ and the
    gravitational mass $M_G$. The solid line corresponds to a constant
    Froude number $\Fr = 2.58$. The inset shows $v^2/g$ \vs\
    $(M_G/\rho w)^{1/2}$, a scaling for which Eqn.~\ref{Fr_def}
    predicts the boundary to be a straight line.  }
  \label{vc1}
\end{figure}
Fig.~\ref{vc1} shows the variation of $v_c$ with the gravitational
mass $M_G$ for the plow.  The frontier between the flat bed and the
appearance of washboard ripples is well described by an $M_G^{1/4}$
scaling, corresponding to a critical Froude number of $\Fr = 2.58$.
This result clearly validates the theory presented in the previous
section, and indicates that the instability is controlled by $\Fr$.
It is interesting to note that the first patterns encountered above
onset are backward traveling ripples for which the plow does not lift
off the surface.  These are discussed in more detail in
section~\ref{sec:backward} below.

\begin{figure}[htbp]
  \rule{1em}{0pt}\textbf{a)}\hfill\mbox{}\\[-1ex]
  \includegraphics[width=70mm,clip]{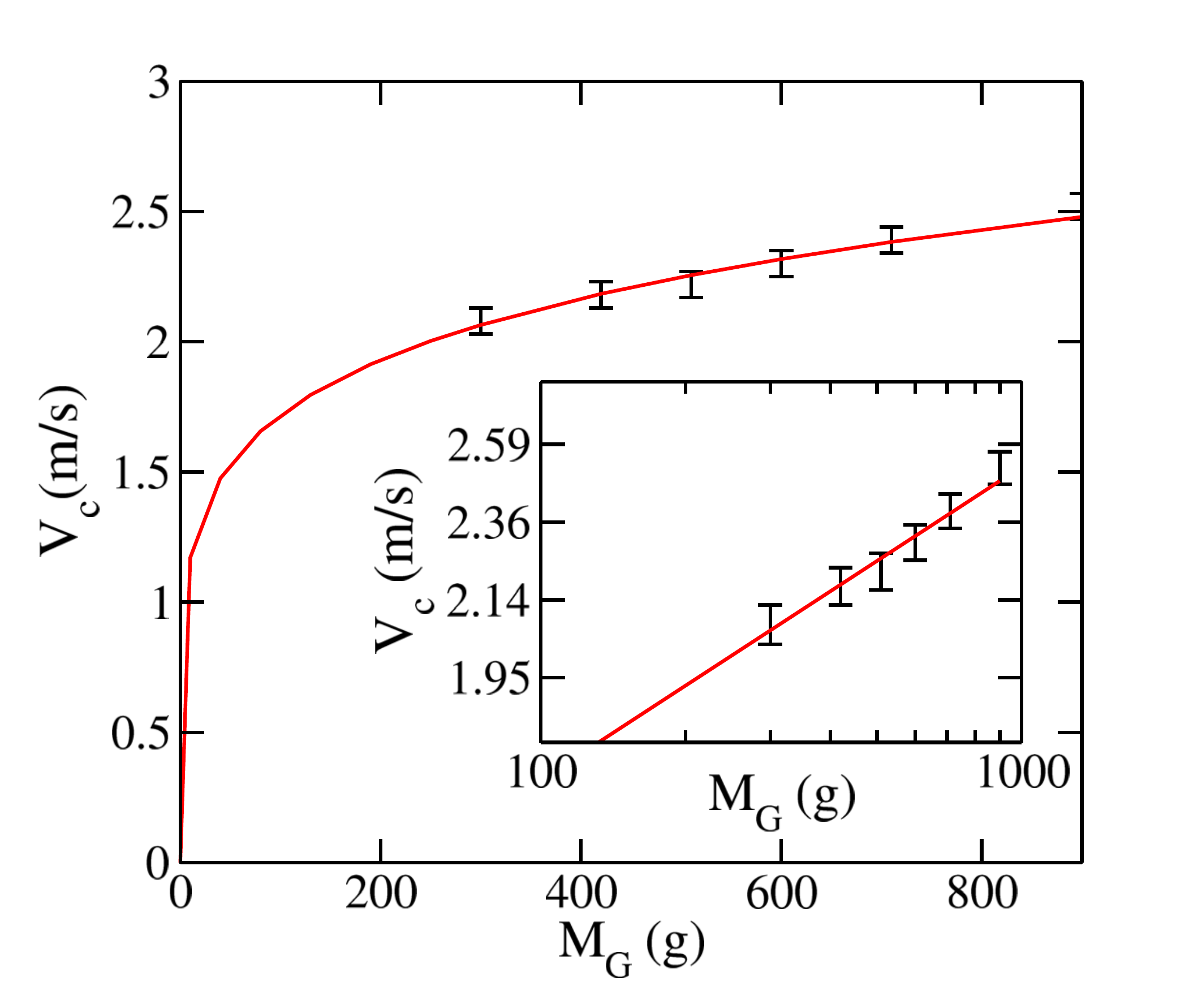}
  \\
  \rule{1em}{0pt}\textbf{b)}\hfill\mbox{}\\[-1ex]
  \includegraphics[width=70mm,clip]{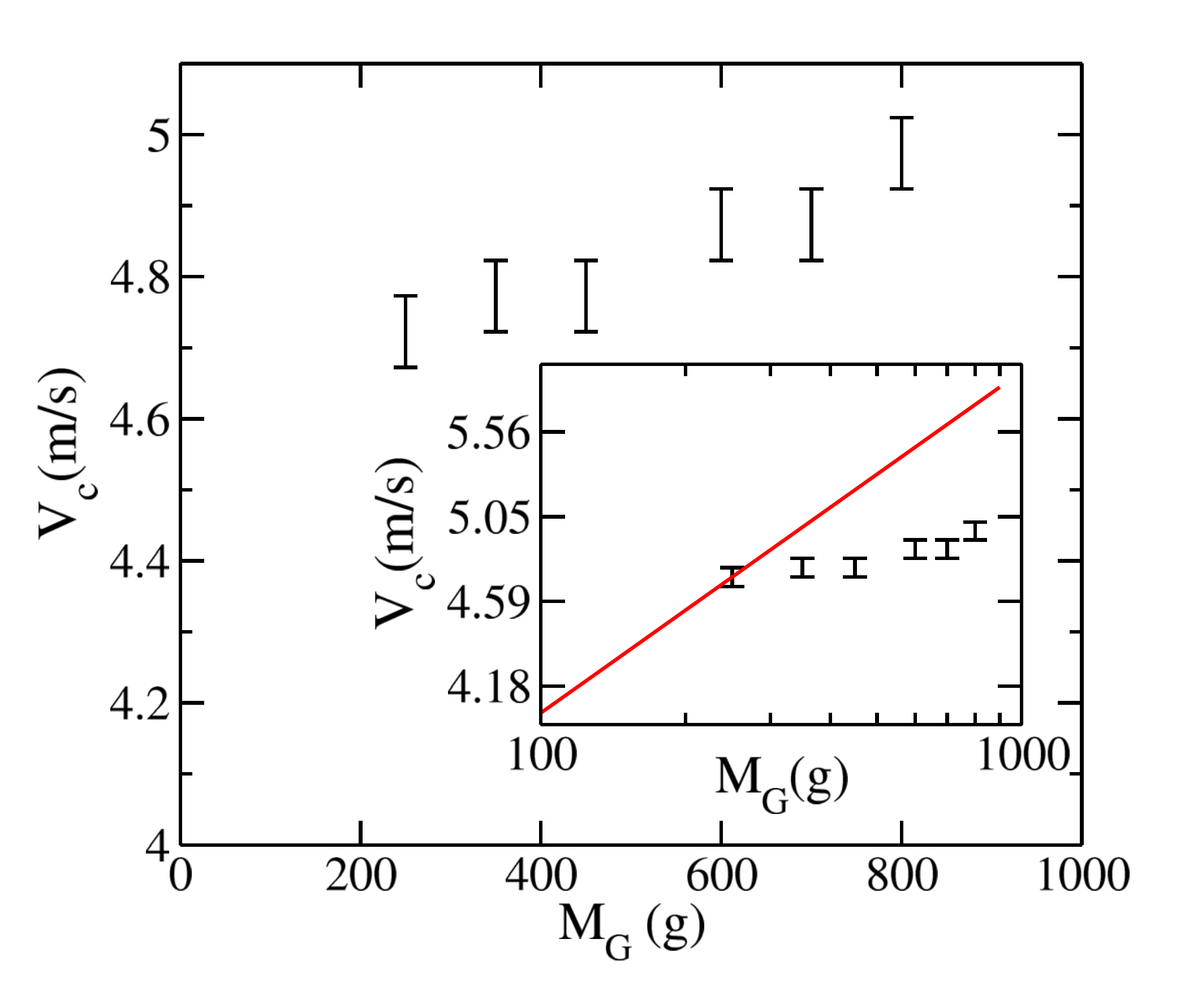}
  \caption{(Color online) The critical speed $v_c$ for the onset of the
    ripple pattern for wheels.  Part (a) shows $v_c$ \vs\ $M_G$ for
    ``blocked", or non-rolling wheels, while (b) shows the same data for
    freely rolling wheels. The solid line corresponds to a constant
    Froude number, according to Eqn.~\ref{vc_wheel}, \ie\  to an
    $M_G^{1/6}$ power law dependence.  It is evident that the predicted
    scaling is not obtained for rolling wheels. }
  \label{fig:blockwheel}
\end{figure}

The scaling theory is less successful for the case of the rolling
wheel; Fig.~\ref{fig:blockwheel} shows $v_c$ \vs\ $M_G$ for the
wheel. To isolate the effects of rolling from the purely geometrical
dependence, we measured $v_c$ both for wheels that were free to roll
and for otherwise identical wheels that were ``blocked" --- \ie\
prevented from rotating.  A blocked wheel behaves much like a
differently shaped plow.

\begin{figure}[htbp]
  \includegraphics{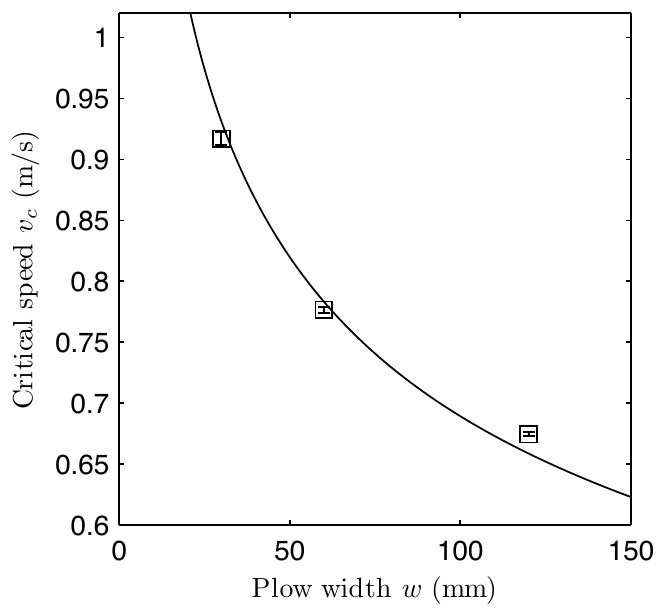}
  \caption{The dependence of the critical velocity $v_c$ on the width
    $w$ of the plow.  The Froude numbers of the three points shown
    were 1.84, 1.86 and 1.99, and hence the critical Froude number
    $\Frc$ remained approximately constant while $w$ was varied. The
    solid line shows a satisfactory fit to the $w^{1/4}$ scaling
    expected from Eqn.~\ref{vc_def}.}
\label{fig:width}
\end{figure}

Fig.~\ref{fig:blockwheel}a shows that the predicted $M_G^{1/6}$
scaling is consistent with the blocked wheel data.  However, experimental limitations
only allowed us to vary $v_c$ over a range of 20\%, so the correctness of 
this scaling is not conclusively demonstrated. 
It is clear however that this scaling fails
for the case of a rolling wheel, as shown in
Fig.~\ref{fig:blockwheel}b. This is perhaps not surprising since
arguments presented in section~\ref{sec:theory} were based only on the
geometry of the wheel and completely ignored its rotation.  Evidently,
the rolling component of the motion affects the surface penetration
properties of the wheel, invalidating the scaling based only on
geometry.  The contact forces for a rolling wheel, and their effect on
the deformation of the granular surface, presumably scale with
additional powers of $R$ and perhaps also depend on the moment of
inertia of the wheel.  In any case, Fig.~\ref{fig:blockwheel} shows
that the gravitational mass dependence of $v_c$ is extremely weak for
wheels, rolling or otherwise.

\subsection{Variation with plow width $w$}
\label{ssec:width}

The scaling argument presented in section~\ref{sec:scaling} depended
on the assumption that the motion was essentially two-dimensional so
that only the two-dimensional density $\rho w$ mattered.  This is
probably the most drastic approximation in the scaling theory.
Figure~\ref{fig:width} shows how $v_c$ scales with $w$ for the plow,
for constant $\alpha$ and $M_G$.  The data is well fit by a $w^{-1/4}$
dependence, as predicted by Eqn.~\ref{vc_def}, and we do find that the
critical value of $\Fr$ remains constant even as $w$ is varied by a
factor of four.  The $w$ scaling can only hold for a relatively small
range of plow widths, however. For very narrow plows, $w \sim h$ and
the two-dimensional scaling will break down.  On the other hand, for
very wide plows $w$ becomes a significant fraction of the radius of
the circular bed and thus there is a significant variation of $v$
across the width.  Figure~\ref{fig:width} spans the intermediate range
of $w$ where a reasonable agreement is obtained.

\begin{figure}[htbp]
  \includegraphics{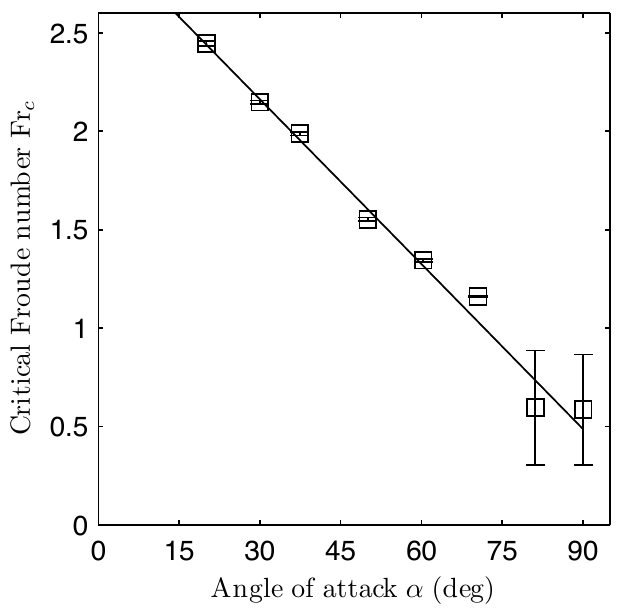}
  \caption{The dependence of the critical Froude number on the angle
    of attack of the plow. The angle $\alpha$ is measured with
    respect to the horizontal (see Fig.~\ref{fig:exp_schematic}). At
    low angles, the transition was very sharp and the error bars lie
    within the symbols.} 
  \label{fig:angle}
\end{figure}

\subsection{Variation with the angles $\alpha$ and $\theta$}
\label{ssec:angle}

The onset of the washboard ripples depended strongly on the angle of
attack of the plow $\alpha$, and on the arm angle $\theta$, as defined
in Fig.~\ref{fig:exp_schematic}. $\alpha$ could be varied, while
keeping all other parameters constant.  $\theta$ depended on the
height of the supporting structure, and could be adjusted
independently of $\alpha$.  Due to experimental limitations, $\theta$
could only be varied over a range of about $15^o$.  The two angles
were in fact somewhat interdependent, as the mean height of the bed
could slowly change over time due to compaction or to sand moving to
other parts of the table.  In such cases, $\alpha$ and $\theta$ both
drifted upward by a few degrees over the course of an experiment.

The simple scaling theory of Eqn.~\ref{vc_def} does not explicitly
predict the $\alpha$ and $\theta$ dependence of the critical speed
$v_c$, which would be expected to show up as a variation of the
critical Froude number \Frc. Fig.~\ref{fig:angle} shows how \Frc\,
depended on $\alpha$.  We found that $v_c$, and hence \Frc, strongly
decreased as $\alpha$ increased. This can be explained by the
observation that the transport of sand along the bed increases when
$\alpha$ is increased. It is not a surprise, therefore, that the
velocity needed to trigger the washboard instability should decrease
with increasing $\alpha$. The case of $\alpha \approx 90^\circ$, a
nearly vertical plow, is somewhat peculiar. At this point, the
transition becomes poorly defined and the amplitude noise near onset
greatly increases.  At $\alpha \approx 90^\circ$, it appears that
other parameters such as the thickness of the plow may play an
important role.

The systematic interdependence of $\alpha$ and $\theta$ makes it quite
difficult to exactly repeat the experimental conditions.  The error
bars in Fig.~\ref{fig:angle} are the precision that the transition can
be identified within a particular experimental run rather than the
somewhat larger error bars that would be obtained from trying to
achieve exactly the same $\theta$ over many runs.

\begin{figure}[htbp]
  \includegraphics{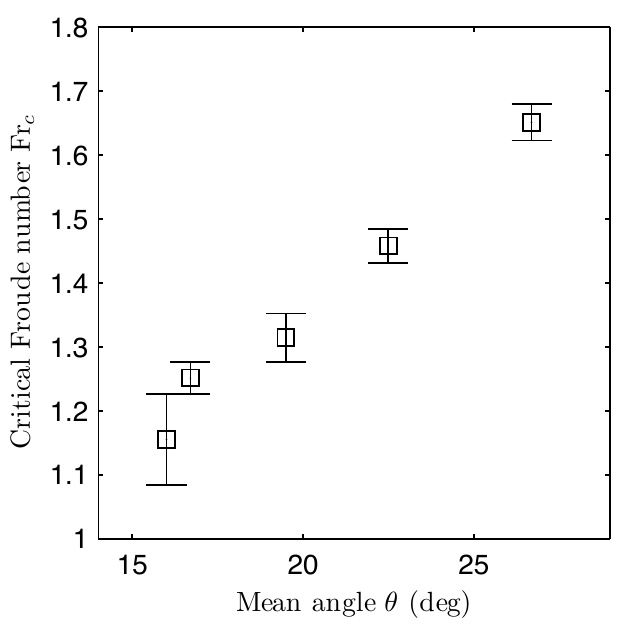}
  \caption{The dependence of the critical Froude number on the mean
    angle $\theta$ that varies with the height of the support.}
  \label{fig:height}
\end{figure}
The critical speed $v_c$ also shows a strong dependence on $\theta$.
Fig.~\ref{fig:height} shows how the critical Froude number scales
with variations in $\theta$, with $\alpha = 45 \pm 3 ^\circ$.  If we
directly apply Eqn.~\ref{eq:z0}, and make a na{\"i}ve model of the
friction forces, we would guess that
\begin{equation}
  v_c \sim  \left[\frac{g^2 M_G}{\rho w} \right]^{1/4}    
  \left(\frac{\cos(\phi-\theta)}{\cos\theta+\mu\sin\theta}\right)^{1/4} ,
\end{equation}
where $\mu$ is the unknown ratio between the normal force $N$ and the
tangential force $T$. For $O(1)$ values of $\mu$ and small $\phi$,
this predicts that the critical Froude number should decrease weakly
with increasing $\theta$. This is the opposite of what is observed.
Clearly, the $\theta$ dependence of the critical Froude number is not
so easily explained and a more detailed study is required to account
for it.

\section{Backward-traveling ripples}
\label{sec:backward}

\begin{figure}[htbp]
  \includegraphics[width=70mm,clip]{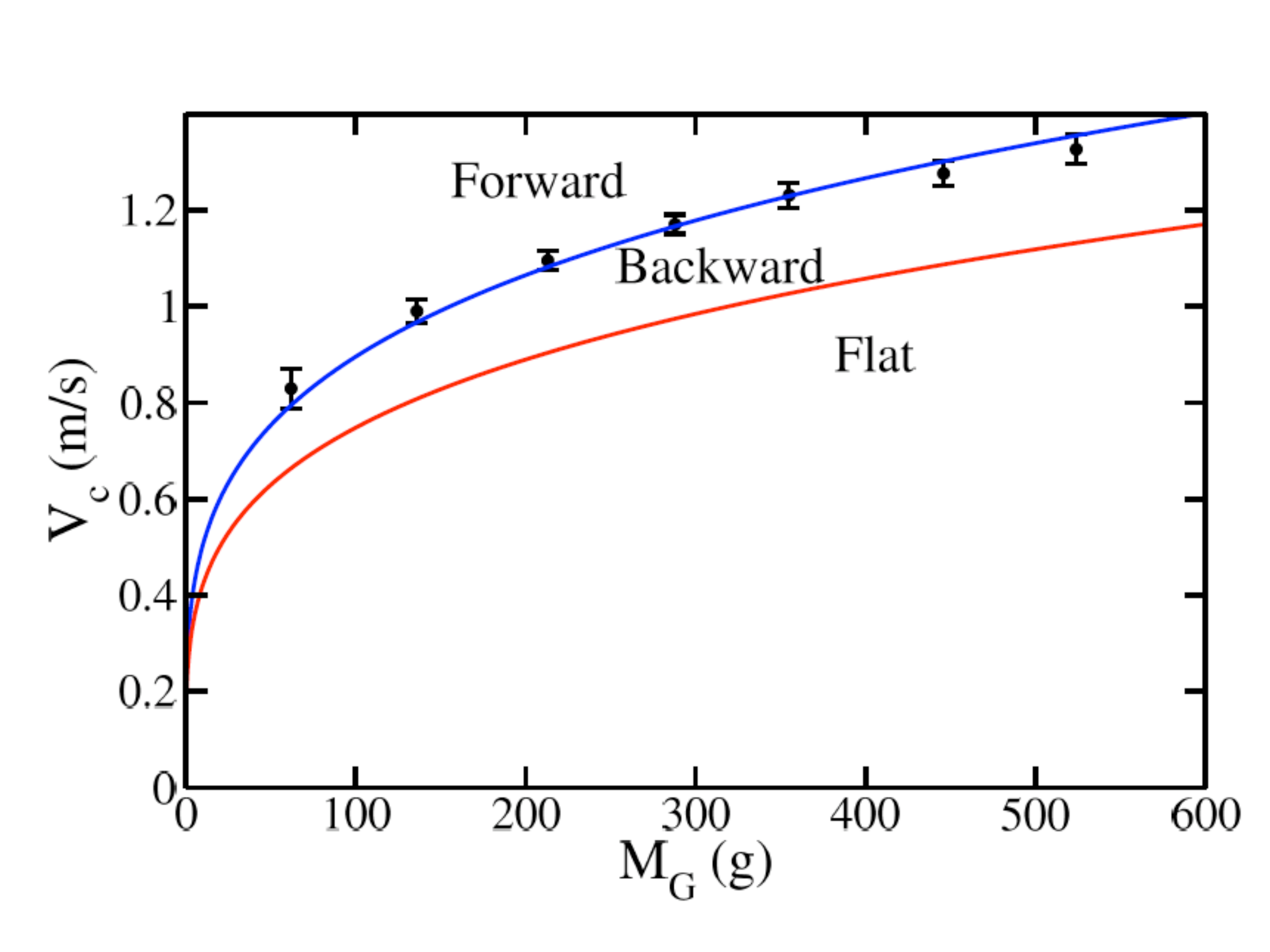}
  \caption{(Color online) The domain of existence and transitions for
    forward and backward traveling ripples using the plow.  The data
    and upper 
    solid line shows the transition between backward and
    forward traveling ripples at $\Fr = 3.80$.  The lower 
    solid line
    shows the onset of backward ripples from the flat bed at $\Fr =
    2.58$, as in Fig.~\ref{vc1}.}
  \label{vt}
\end{figure}

In all experiments using wheels, including our previous
studies~\cite{taberlet07:washboard}, the ripples that appear just
above onset were observed to travel in the direction of driving.  This
agrees with most, but not all, observations in previous engineering
studies~\cite{heath80:washreview}.  However, ripples produced by a
plow just above onset always travel in the reverse direction, \ie\
opposite to the direction of driving, as shown in
Fig.~\ref{fig:backwards}.  The backward travel reverses at higher
driving speeds, but there is no smooth transition from backward to
forward ripples; there is no speed where the ripples are stationary.
If the speed is rapidly stepped from below to well above $v_c$, as in
Fig.~\ref{fig:init}b, the threshold speed above which backward ripples
are observed in the steady state appears to scale in the same way as
$v_c$, as shown in Fig.~\ref{vt}.  Backward traveling ripples only
exist for a narrow range of parameter values just above onset.

Another distinguishing feature of backward traveling ripples is that
ripple merger events, which are common near onset for forward
traveling ripples, generally do not occur. This is because the
backward ripples appear at the same wavelength as in their
amplitude-saturated state, whereas there is a strong relationship
between the wavelength and amplitude for the forward ripples. This is
related to another difference: that the plow does not leave the
granular surface during backward ripple motion, while such ballistic
motion is always present for forward traveling ripples using the plow.

\begin{figure}[htbp]
  \includegraphics[width=70mm,clip]{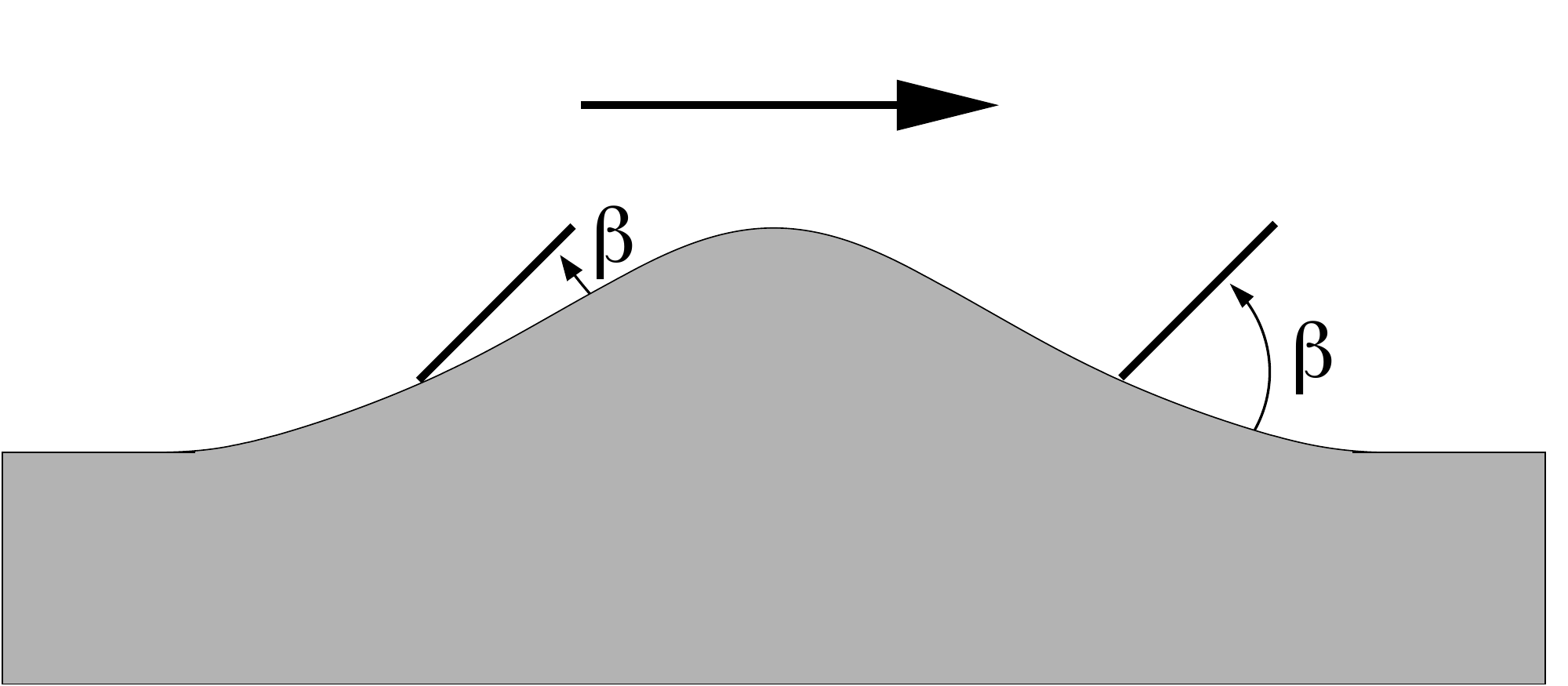}
  \caption{Schematic mechanism for the formation of backward traveling
    ripples.  The arrow shows the direction of motion of the plow. If
    the rate of erosion by the plow increases with the angle $\beta$
    with the surface, then material is preferentially removed from the
    downward face of the ripple and deposited on the upward face of
    the following one, giving rise to a net backward ripple motion.
    During this process, the plow does not leave the surface. }
 \label{sketch_back}
\end{figure}

The formation of backward traveling ripples is fundamentally different
from the inertial mechanism of forward traveling ripples.  Ripples can
move backward even though the sand is only transported forward.  The
surface flux of grains in the driving direction depends not only on
the normal force, but also on the geometry of the contact region.  It
is plausible to suppose that the surface flux increases with the angle
$\beta$ between the plow and the granular surface.  Initially, $\beta
= \alpha$, the fixed angle of attack of the plow, but as the ripples
develop, $\beta$ is modulated by the ripples as the plow remains in
contact with the surface, as shown in Fig.~\ref{sketch_back}.  The
result of this modulation is that $\beta$, and hence the surface flux,
is smaller on the upward face of the ripple than on the downward one.
The upward face is therefore less eroded, while the sand that is
preferentially eroded from the downward face is deposited on the
upward face of the following ripple.  Thus, the net motion of the
ripple is backward, while the grains are still transported forward.  A
similar mechanism has been proposed for the uphill motion of moguls on
ski slopes~\cite{mogul_motion}.  

Conversely, in the case of forward traveling ripples, inertial effects
dominate and the plow leaves the surface and goes into ballistic
motion near the crest of the ripple.  The erosion is therefore all on
the upward face, while the downward one forms a slip face which is
added to by small avalanches consisting of material transported from
the upward face.  Thus, the ripples in the ballistic regime move
forward, in the driving direction.

\section{More complex states}
\label{sec:zoo}
\begin{figure*}[htbp]
  \includegraphics{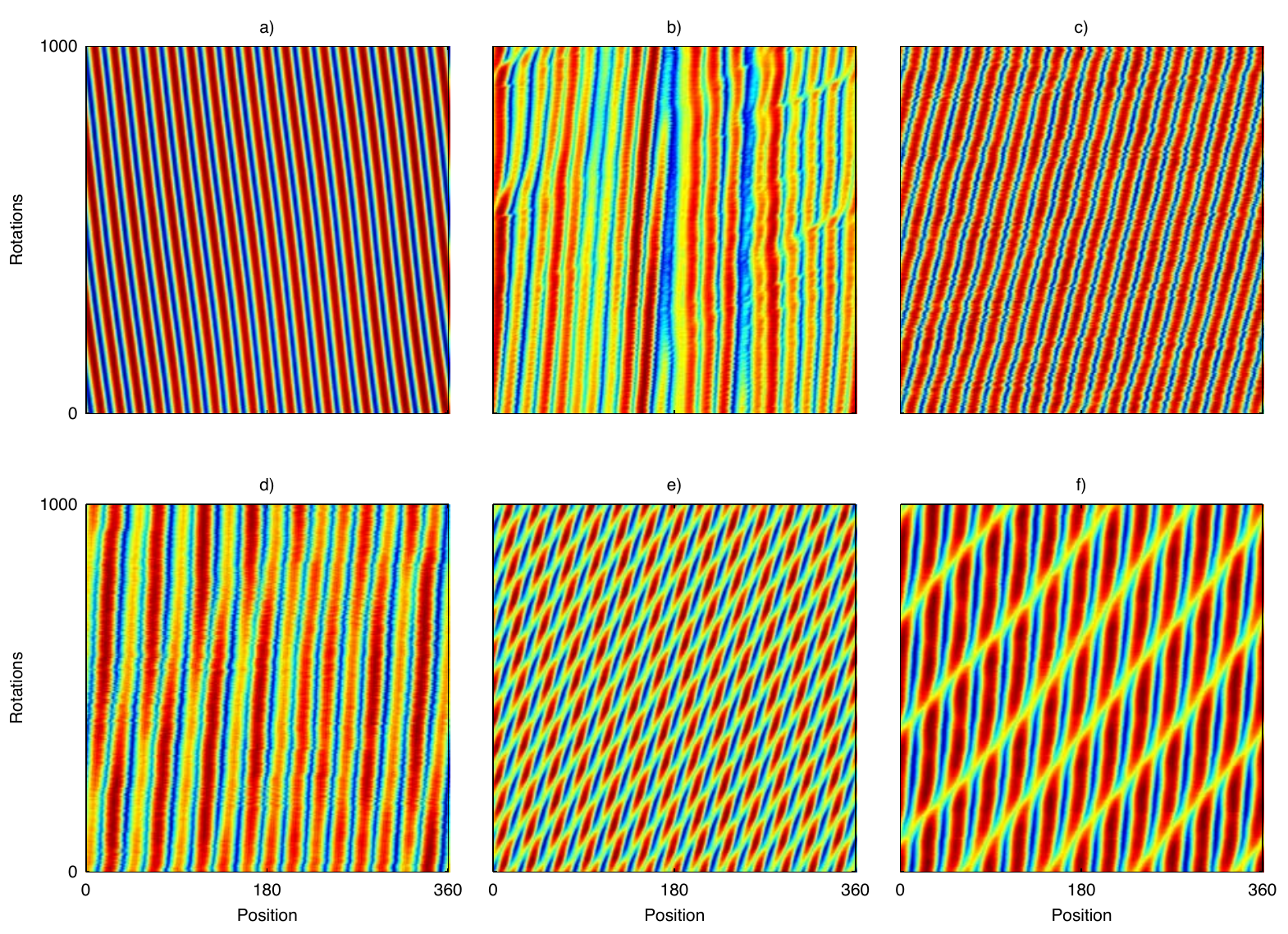}
  \caption{The zoo of states for the plow and wheel. Each figure
    contains 1000 revolutions at constant speed. The range of ripple
    heights is normalized for each figure with red representing higher
    a) Plow at \speed{0.737}. b) Plow at \speed{0.827}. c) Plow at
    \speed{0.964}. d) Plow at \speed{1.001}. e) Wheel at
    \speed{1.166}. f) Wheel at \speed{1.325}}.
  \label{fig:zoo}
\end{figure*}

In the preceding sections, we have focused on the scaling of the
critical speed $v_c$ for the onset of the ripple pattern.  While a
complete study of the states above onset is beyond the scope of this
paper, it is interesting to survey some of the complex states that are
encountered above $v_c$.  Fig.~\ref{fig:zoo} shows a selection of such
fully developed ripple states.

Other than the flat bed, the simplest states are pure traveling
ripples, which are well described by single Fourier modes.  For
wheels, these always travel forward, while for the plow they travel
backward near onset, as in Fig.~\ref{fig:zoo}a, and forward for larger
speeds, as in Fig.~\ref{fig:zoo}c.  As the driving speed $v$ of the
plow is increased, the backward traveling ripple speed decreases, but
is never zero.  Increasing $v$ results in a low-dimensional chaotic
state, as in Fig.~\ref{fig:zoo}b, in which several Fourier modes are
excited.  Further increasing $v$, we enter a regime of steady forward
traveling ripples, but here the plow spends much of the time in a
free, ballistic trajectory between impacts. If these impacts are not
periodic around the table, there is some jitter between the phase of
the ripples and that of the table; this is visible in
Fig.~\ref{fig:zoo}c.  At still higher speeds of the plow, a second
chaotic regime is reached, as shown in Fig.~\ref{fig:zoo}d.

The phenomenology for wheels differs in detail, but is similarly
complex.  For example, Fig.~\ref{fig:zoo}e and~f show wheel-driven
ripple states consisting of two significant Fourier modes which are
not simply related.  The two ripple modes propagate with different
speeds, producing a beating effect which persists indefinitely.  All
of these complex states for plows and wheels are reminiscent of the
multiple dynamical states of impact oscillators~\cite{bishop94:impact}
in their strongly nonlinear regime.

\section{Conclusion}
\label{sec:conclusion}

We have studied the rippling instability of a flat granular surface
under the action of a moving source of stress.  We considered both the
classic case of a rolling wheel producing ``washboard road" and the
simpler case of a flat plow blade.  In both cases, we focused on the
region near the onset speed $v_c$.  We developed a scaling theory for
$v_c$ based on the idea that the onset of the instability represents a
crossover between conservative, dynamic forces proportional to $v^2$
and dissipative, static forces which are independent of $v$.  We found
that $v_c$ did scale with the gravitational mass as $M_G$ as predicted
for the plow and for non-rolling wheels.  These have different
scalings due to differences in their shape.  However, the predicted
mass scaling was not found for rolling wheels, suggesting that a more
detailed theory of the surface penetration of a rolling wheel is
required.  We then considered the dependence of $v_c$ on various other
geometric parameters.  

Finally, we surveyed some of the complex ripple
dynamics that are observed above onset, which include both forward and
backward traveling ripples, chaotic states and states with traveling
defects.  
%
%
Clearly, understanding the above-onset regime of finite amplitude ripples and their
interactions will require a fully nonlinear theory of the instability.  Such a theory would need to go
well beyond the simple scaling of the onset speed $v_c$ that we have discussed here.

Throughout this work, we have used the angular position of the arm to
trace the motion of the wheel or plow.  To completely close the
feedback loop between the moving source of stress and the state of the
granular bed, we would also need to measure the shape of the granular
surface and the rotational state of the wheel.  This would allow us to
determine the full acceleration of the wheel or plow, and hence the
forces on it, as well as the dynamic response of the granular bed. The
bed shape could be determined, for example, by a scanning laser
profilometer~\cite{laser_triangulator}.  Another, somewhat
complementary avenue to the same kind of information is soft sphere
molecular dynamics simulation~\cite{taberlet07:washboard} in 2D or 3D.
With this data, a realistic dynamical model of the wheel-bed
interaction could be developed, leading to a more complete continuum
theory of the washboard instability.

\begin{acknowledgments}
 This work was supported by the George Batchelor Laboratory,
 University of Cambridge, the Swiss National Science Foundation, the Natural
 Science and Engineering Research Council (NSERC) of Canada, the
 Isaac Newton Trust and the Engineering and Physical Sciences
 Research Council (UK). Many thanks to Stuart Dalziel, director of
 the Batchelor laboratory.
\end{acknowledgments}


\begin{thebibliography}{25}

\bibitem{heath80:washreview} W. Heath and R. Robinson, {\it Transport
    and Road Research Laboratory}, Suppl. Report No. 610 (1980).

\bibitem{mather62:washboard1} K. B. Mather, {\it Civ. Eng. Pub. Works
    Rev.}, {\bf 57}, 617 (1963), and {\it Civ. Eng. Pub. Works Rev.},
  {\bf 57}, 781 (1963).

\bibitem{shoop06:washboard} S. Shoop, R. Haehnel, V. Janoo, D. Harjes
  and R.  Liston, {\it J. Geotech. Geoenviron. Eng.}, \textbf{132},
  852 (2006).

\bibitem{riley73:washboard} J. G. Riley and R. B. Furry, {\it Highway
    Research Record}, {\bf 438}, 54 (1973).

\bibitem{riley71:washboard} J. G. Riley, Ph.D thesis, Cornell
  University (1971).

\bibitem{grau91:washboard} R. W. Grau and L. B. Della-Moretta,
  \emph{Trans. Res. Rec.} \textbf{1291}, Vol. 2, 313 (1991).

\bibitem{misoi89:washboard1} G. K. Misoi and R. M. Carson, {\it Proc.
    Instn. Mech. Engrs.}, {\bf 203}, 205 (1989), and {\it Proc. Instn.
    Mech. Engrs.}, {\bf 203}, 215 (1989).

\bibitem{taberlet07:washboard} N. Taberlet, S. W. Morris and J. N.
  McElwaine, {\it Phys. Rev. Let.}, {\bf 99}, 068003 (2007).

\bibitem{cross04:pattern} M. C. Cross and H. Greenside, {\it Pattern
    Formation and Dynamics in Nonequilibrium Systems}, Cambridge
  University Press, 2009.

\bibitem{mays00:washboard} D. C. Mays and B. A. Faybishenko, {\it
    Complexity}, {\bf 5}, 51 (2000).

\bibitem{both01:washboard} J. A. Both, D. C. Hong and D. A. Kurtze,
  {\it Physica A}, {\bf 301}, 545 (2001).

\bibitem{bagnold1941} R. A. Bagnold, {\it The Physics of Blown Sand
    and Desert Dunes}, Methuen, London 1941.

\bibitem{engelund82:ripples} F. Engelund and J. Fredsoe, {\it Ann.
    Rev. Fluid Mech.}, {\bf 14}, 13 (1982).

\bibitem{sato02:rails} Y. Sato, A. Matsumoto and K. Knothe, {\it
    Wear}, {\bf 253}, 130 (2002).

\bibitem{egger02:moguls} J. Egger, {\it Physica D}, {\bf 165}, 127
  (2002).

\bibitem{clanet04:skipping} C. Clanet, F. Hersen and L. Bocquet, {\it
    Nature}, {\bf 427}, 29 (2004).

\bibitem{dai04:washboard} Q. Dai, F. Hendriks and B. Marchon, {\it J.
    Appl. Phys.}, {\bf 96} 696 (2004).

\bibitem{pit01:mogul} R. Pit, B. Marchon, S. Meeks and V. Velidandla,
  {\it J. Trib. Lett.}, {\bf 10} 133 (2001).

\bibitem{chin94:impact} W. Chin, E. Ott, H. E. Nusse and C. Grebogi,
  {\it Phys. Rev. E}, {\bf 50} 4427 (1994).

\bibitem{bishop94:impact} S. R. Bishop, {\it Phil. Trans. R. Soc.
    Lond. A}, {\bf 347}, 347 (1994).

\bibitem{magnetic_sensor} POSIROT from ASM limited. Part No:
  PRAS1-30-U2-CW-M12 modified to operate at 2\,kHz.
                  
\bibitem{froude} The hydrodynamic Froude number is given either by
  $v^2/gL$, or sometimes by the square root of this quantity.  It is
  usually physically explained as being the ratio of a speed $v$ and
  the wave speed of a surface gravity wave of wavelength $L$, all
  squared.  Of course, no gravity waves are involved in the formation
  of the washboard ripple pattern.
                  
\bibitem{mogul_motion} David Bahr, private communication.
   \htmladdnormallink{[Link]}{http://academic.regis.edu/dbahr/GeneralPages/Research/MogulResearch.htm}.
  
\bibitem{laser_triangulator} LLT2800-100 laser triangulator from
  Micro-Epsilon Ltd.
\end{thebibliography}
\end{document}